\documentclass[12pt]{article}
\usepackage{times}
\usepackage{geometry}
\usepackage[utf8]{inputenc}
\usepackage{multicol}
\usepackage{ragged2e}
\usepackage{pifont}

\usepackage{mathptmx} 
\usepackage{graphicx}
\usepackage{amssymb}
\usepackage{epstopdf}
\usepackage{enumitem}
\usepackage{fancyhdr}
\usepackage{amsmath}
\usepackage[font={footnotesize, it},labelfont={footnotesize, bf}]{caption}
\usepackage[font=small,skip=0pt]{caption}
\usepackage{epstopdf}
\usepackage{hyperref}
\usepackage{natbib}
\usepackage{sectsty}
\sectionfont{\fontsize{13}{15}\selectfont}

\newcommand{\farcs}{\mbox{\ensuremath{.\!\!^{\prime\prime}}}}
\newcommand{\arcsec}{\ensuremath{^{\prime\prime}}}              

\begin{document}

\thispagestyle{empty}

\raggedright
\huge
Astro2020 Science White Paper \linebreak

Probing Unseen Planet Populations with Resolved Debris Disk Structures \linebreak
\normalsize

\noindent \textbf{Thematic Areas:} \hspace*{60pt} $\boxtimes$ Planetary Systems \hspace*{10pt} $\boxtimes$ Star and Planet Formation \hspace*{20pt}\linebreak
$\square$ Formation and Evolution of Compact Objects \hspace*{31pt} $\square$ Cosmology and Fundamental Physics \linebreak
  $\square$  Stars and Stellar Evolution \hspace*{1pt} $\square$ Resolved Stellar Populations and their Environments \hspace*{40pt} \linebreak
  $\square$    Galaxy Evolution   \hspace*{45pt} $\square$             Multi-Messenger Astronomy and Astrophysics \hspace*{65pt} \linebreak
  
\textbf{Principal Author:}

Name: Kate Y. L. Su
 \linebreak					
Institution:  Steward Observatory, University of Arizona, Tucson, AZ, USA  
 \linebreak 
Email: ksu@as.arizona.edu \ \ \ \ Phone:  520-621-3445 \\
\medskip
\textbf{Co-authors:}
Nick Ballering, Steve Ertel, Andras Gaspar, Grant Kennedy, David Leisawitz, Meredith MacGregor, Brenda Matthews, Amaya Moro-Martin, George Rieke, Jacob White, David Wilner, Mark Wyatt (other co-signers in the reference) \\
\medskip
\textbf{Abstract  (optional):}\\

Debris disks emerge when larger objects stir remnant planetesimal belts, resulting in cascades of collisions that break minor bodies (asteroids and comets) down into dust to be heated by the star, making them detectable over the whole duration of a stellar lifetime. Planets imprint their signatures on the configuration of these planetesimals: the location of the planetesimal belts is governed by where the planets form, locate and their migration history. The structure of the minor bodies in our Solar System contains clues of the past giant planet migrations and reveals the current influence of the giant planets. Thousands of exoplanets have been found with many widely different from the ones in our own system. Despite the success, systems with planets in wide orbits analogous to those of Jupiter and Saturn, in the critical first several hundred million years of evolution, are virtually unexplored. Where are the low-mass planets that are hidden from our exoplanet detection techniques? Is our Solar System's planetary architecture unique? High-fidelity debris disk images offer an effective method to answer these questions. We can use them to study the formation and evolution of low-mass planets from youth to the age of the Solar System, providing snapshots of the complex processes and valuable insights into the formation and migration history of giant planets at wide orbits. This white paper focuses on resolving debris structures in thermal emission that is applicable to a large unbiased sample. We summarize the properties of the known debris disks and assess the feasibility of resolving them within our current and future infrared and millimeter facilities by adopting uniform criteria. {\it JWST} and the 9-m {\it Origins} Space Telescope are the most promising missions in the coming decades to resolve almost half of the known disks at high fidelity. We emphasize the need to resolve disks at multiple wavelengths, particularly in the millimeter range. Resolved debris structures at multiple wavelengths and at all stages of evolution would reveal the properties of unseen planet populations, enabling a unique demographic study of overall planet formation and evolution.

\justifying
\frenchspacing

\pagebreak
\setcounter{page}{1}

\section{Background and Motivation}
\vspace{-3mm}

\begin{figure*}[h!]
    \includegraphics[width=1.0\textwidth]{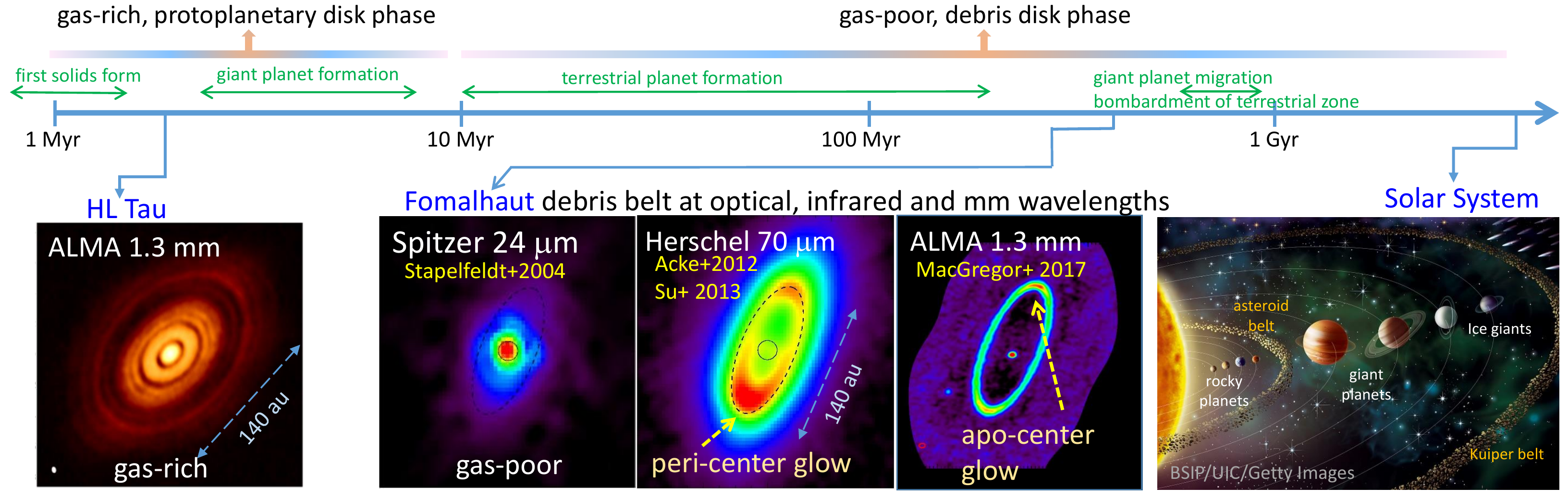}   
  \caption{\small The timeline of the formation and evolution of a planetary system with marks (green lines) indicating important phases in our Solar System. Snapshots of disk images such as HL Tau (far left panel, ALMA collaboration) and Fomalhaut (right three panels) bare the imprints of unseen planets that can be used to reveal the underlying planetary configuration. Snapshots of disk images at different evolutionary stages provide insights into the complex processes on how our Solar System (far right panel) came to be.}
  \label{fig:hltau_fom}
\end{figure*}

In roughly two decades, we have progressed from possessing evidence for only one planetary system, our own, to thousands confirmed today. However, our most successful methods for the detection of exoplanets are mostly biased toward the inner zone of mature systems. Most known exoplanetary systems are radically different from the Solar System in ways that make them unlikely abodes for life. Having giant or ice-giant planets at large orbital distances not only fosters the formation of terrestrial planets (Raymond et al.\ 2012) but also serves as a shield from the influx of small bodies to habitable terrestrial planets. Systems with planets in wide orbits analogous to those of Jupiter and Saturn, in the critical first few hundred million years of evolution, are virtually unexplored. Studying debris disks offers an alternative method to characterize planetary systems and their evolution (Wyatt 2008; Krivov 2010; Matthews et al.\ 2014; Hughes et al.\ 2018). These disks are composed of dust grains ranging from $\sim\mu$m to mm-sized grains continually replenished by sublimation and collisions of planetesimals as the byproduct of planet formation. Debris disks, identified as infrared excesses around stars, trace a pattern of development thought to be similar to that of the Solar System: (1) a peak in inner debris disk activity at 10--30 Myr, when terrestrial planets are being built; (2) thereafter a decay for about ~1 Gyr, the expected time dependence for collisional cascades, and matching the time scale of inward bombardments to the terrestrial region; and (3) occasional major collisions such as the one that led to the formation of our Moon and the Pluto-Charon system. The orbits of the planetesimals are inevitably perturbed by the presence of planets with masses down to sub-Earth mass. Hence, the debris structure can reveal hidden planets in a planetary system at all stages of planetary system evolution. Each of the resolved disk images provides a snapshot in the complex architectural picture of planetary system evolution (Fig.\ref{fig:hltau_fom}), complementing the techniques of exoplanet detection.

\vspace{-3mm}

\vspace{-8mm}
\section{What Resolved Disk Images Probe -- Unseen Planet Populations}
\vspace{-3mm}

\begin{figure*}[h!]
    \centering
    \includegraphics[width=1.0\textwidth]{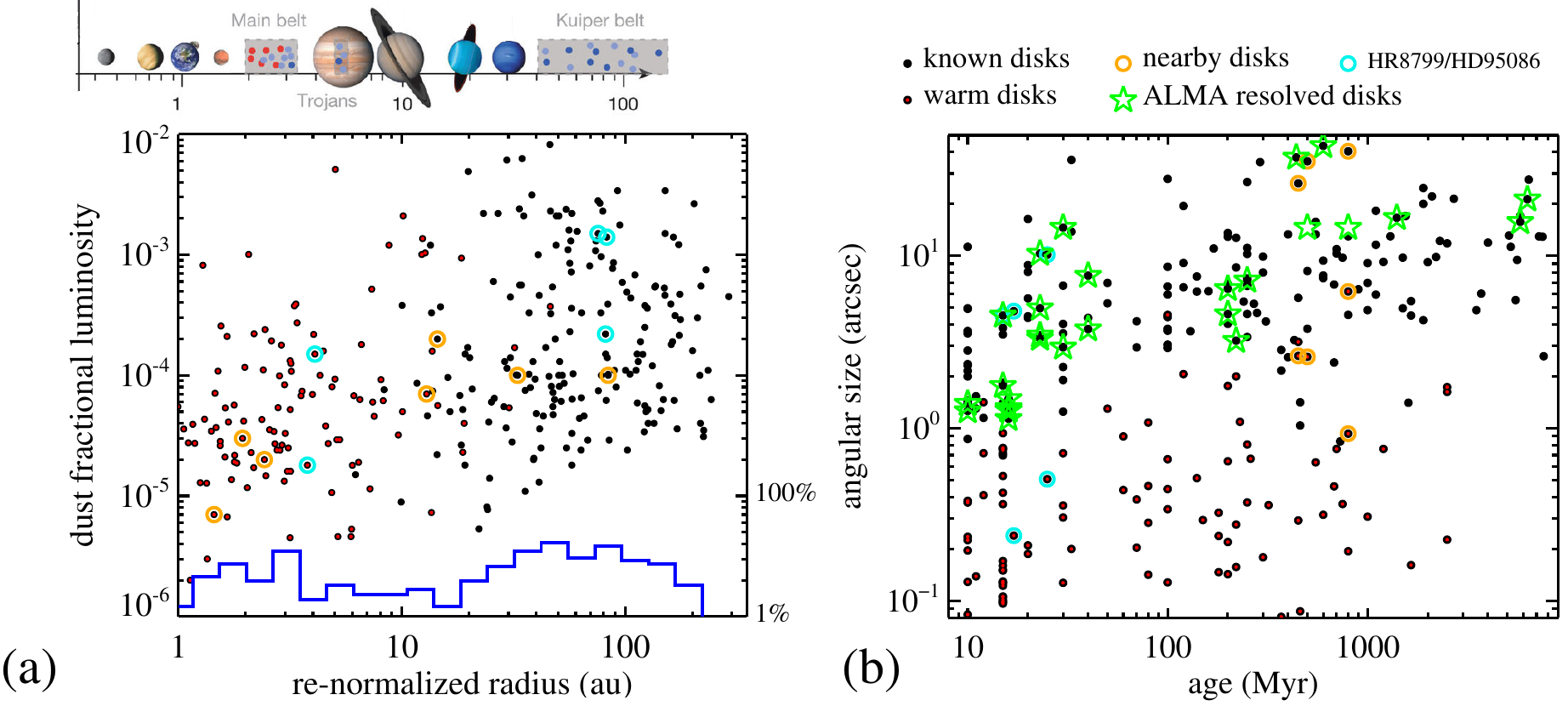}
    \caption{\small Properties of the known debris systems in the context of our own Solar System. (a) The distribution of the debris belt radii (x-axis, re-normalized to $\sqrt{L_{\ast}}$ so that the snow lines around different types of stars are aligned as indicated by the top solar system sketch) and the amount of debris (y-axis), expressed as the infrared dust fractional luminosity. The blue line is the size histogram in logarithmic bins with the scale displayed on the right-y axis. (b) The age vs. angular size (diameter) distribution in the debris sample suggests a wide diversity in the sizes of planetesimal belts. Each of the systems represents a snapshot in the evolution of a planetary system. We note that 20\% of the systems  have very uncertain ages, therefore, they are not shown in the right panel. 
}
    \label{fig:knowndebris}
\end{figure*}

Thanks to the sensitive infrared surveys provided by modern space telescopes ({\it Spitzer}, {\it Herschel}, and {\it WISE}) we now have identified thousands of infrared excesses around mature stars with $\sim$400 of them within 100 pc (see Fig.\ \ref{fig:knowndebris} for the general properties of the known debris disks). However, most of the debris disks are not spatially resolved, with only broadband photometry (and some with mid-infrared spectroscopy) defining their spectral energy distributions (SEDs). For low levels of excess emission (a few \% of the stellar photosphere), it is difficult to infer disk properties without resolving the disk from the star because the star dominates the noise in unresolved photometry and stellar activity can mimic excess emission. Resolving the excess emission from the star is the only robust way to detect and characterize such faint emission (see one of the discovery science cases, {\it Frequency of True Kuiper-belt Analogs}, in the {\it Origins} final report). \textit{\textbf{Is the size of our Solar System typical among planetary systems? Does the Solar System's size stem from its natal disk, or does its current size reflect its evolution?}}
One of the immediate results of resolved disk images is to have a complete census on the outermost edge of planetary systems.  Debris in the outermost regions, the Kuiper-belt analogs, has emission that peaks in the far-infrared where the stellar contribution is negligible, making the detection of debris emission relatively easy. \textit{A statistical sample of resolved disk images at all evolutionary stages can immediately answer these questions.} 

\vspace{-2mm}
\vspace{-2mm}

The minor bodies in our Solar System are arranged in a structure sculpted by the planets over the course of 4.5 Gyr of evolution. The structure contains clues of the past giant planet migrations (as suggested by the NICE and Grand Tack models) and reveals the current influence of the giant planets (Jupiter's role in the structure of the asteroid belt and Neptune's role in that of the Kuiper belt). Any debris system with similar structures may be signaling the presence of planets. One of the best examples is the HR 8799 system. From the marginally resolved images and detailed SED modeling, two major dust components are inferred: cold Kuiper-belt and warm asteroid-belt analogs (Su et al.\ 2009). The four directly imaged massive planets lie between them, as expected if orbital resonances maintain the debris disk structure. Resonance structures are not limited to massive planets -- our terrestrial planets are known to create asymmetric features in the circumsolar ring (Reach et al.\ 2010, Stenborg et al.\ 2018). Resolved debris images provide constraints on unseen planets through their interactions with the disks (details see reviews by Wyatt 2008 and Hughes et al.\ 2018), which manifest as variations in debris width and eccentricity, creating offsets and asymmetries in debris belts. By resolving the structure of debris disks around stars of different ages, we would learn the planetary configurations at different times. Doing so with a large sample, we can explore the full evolution of planetary systems and determine whether our Solar System's architecture (terrestrial planets, asteroid belt, giant planets and ice giants and Kuiper belt, right panel of Fig.\ \ref{fig:hltau_fom}) is unique or not, putting our Solar System into context.

Because the debris is mostly generated by collisional cascades of planetesimals, a wide range of particle sizes is present in debris belts, with a typical power-law size distribution. Depending on its stellocentric distance, the emission of the debris peaks at different wavelengths: $\sim$25 $\mu$m for the warm asteroid-like belts and $\sim$60 $\mu$m for the cold Kuiper-belt analogs. Hence, detecting the same debris belt on the Wien or Rayleigh–Jeans sides of the emission requires higher sensitivity. Small grains are further influenced by non-gravitational forces; and their dynamics are altered so they do not necessarily trace the large planetesimals. Observations in different wavebands are sensitive to different dust sizes. Multi-wavelength imaging of debris disks can directly probe the particle size distribution (i.e., signs of dynamical mixing), and further provide stringent constraints on the unseen planets and dust dynamics (Fig.\ \ref{fig:multiwav_resolution}a, Wyatt 2006; Ertel et al.\ 2012). {\it HST} and ground-based 10-m high contrast imaging facilities provide sub-arcsec resolution, and have resolved a few dozens of debris disks. However, scattered light measurements are biased strongly toward young systems with high density zones, preferential for close to edge-on geometry. Resolving unbiased samples requires detection in the thermal emission. {\it Spitzer's} 24 $\mu$m and {\it Herschel's} 70 $\mu$m channels provided the best resolution (6\arcsec) in the past. The cold disks in Vega, Fomalhaut and $\epsilon$ Eri (nearby systems within 10 pc) are resolved by factors of 4, 6 and 6 beam sizes, respectively, enabling the discovery of complex disk structures (Su et al.\ 2013).  Information extracted from marginally resolved images (only a few dozen systems were resolved by {\it Herschel}) is limited, preventing us to fully explore the potential of using debris disks to probe unseen planet populations.   
\textbf{\textit{Where are the low-mass planets that are currently hidden from our exoplanet detection techniques? Is our Solar System's planetary architecture unique?}} \textit{Resolved debris structures at multi-wavelengths and at all stages of evolution would reveal the properties of unseen planet populations, enabling a unique demographic study of overall planet formation and evolution.}

\vspace{-5mm}
\section{Future Possibilities: Resolving Large Disk Samples}
\vspace{-3mm}

To assess the feasibility of resolving debris structures within our current and future capabilities, the first step is to estimate the expected surface brightness (S.B.) at a given wavelength using the measured flux and estimated size of the disks. We have collected a sample of $\sim$400 systems from the literature that have infrared spectroscopic and photometric measurements, adequate to perform a statistical study. In most cases, the disk excess can be described as a modified blackbody with the SED derived dust temperature and an empirical power-law of $F_{\nu}\sim\nu^{\alpha_{mm}}$ and $\alpha_{mm}\sim2.6$ to extrapolate flux to mm and cm wavelengths (MacGregor et al.\ 2016). Calibrated on a few dozens of marginally resolved disks, the disk sizes can be estimated using the relations from (1) Pawellek et al.\ 2014 and (2) Matra et al.\ 2018. For systems where the measured SEDs require two temperatures, the disk components are characterized as the warm (dust temperatures $\gtrsim$100 K) and cold components. To directly compare to the solar system and the associated snowline locations for all different spectral types, Fig. \ref{fig:knowndebris}a shows the re-normalized disk radius distribution (scaled by $\sqrt{L_{\ast}}$) while Fig. \ref{fig:knowndebris}b shows the angular sizes vs. age distribution.

\begin{figure*}[h!]
    \centering
    \includegraphics[width=1.0\textwidth]{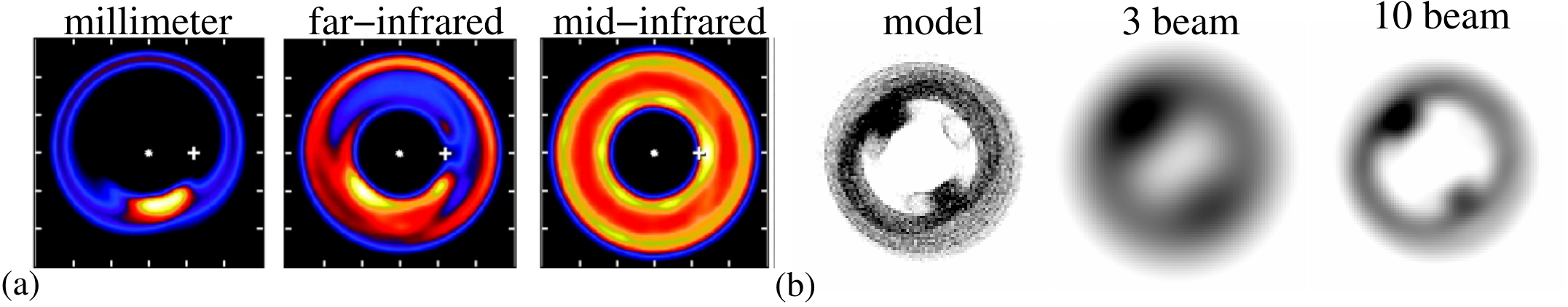}
    \caption{\small (a) Wavelength-dependent disk structures induced by the same planet (plus sign), illustrating the need for multi-wavelength observations to put stringent constraints on the unseen planet (figure extracted from Wyatt 2006). (b) Illustrations of the resolvability (defined as the ratio between the disk diameter and the observed beam) to probe disk structures. Resolving by $\gtrsim$3 beams can pin point the location of the belt and detect modest asymmetry while the images resolving at 10 beams are very diagnostic to resonance structures that are sensitive to the mass and location of a (unseen) perturbing planet.}
    \label{fig:multiwav_resolution}
\end{figure*}

\begin{figure*}
    \centering
    \includegraphics[width=1.0\textwidth]{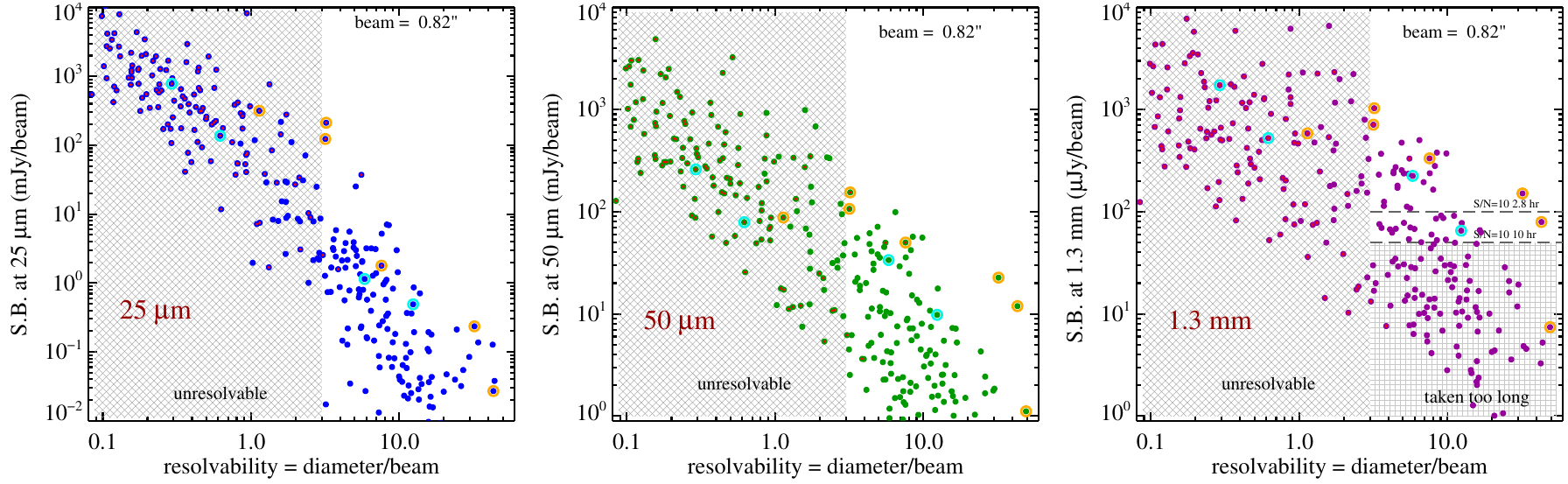}
    \caption{\small The expected surface brightness (S.B.) vs. the resolvability using a beam of 0\farcs82 at 24 $\mu$m (left), 70 $\mu$m (middle) and 1.3 mm (right). Symbols used are the same as in Figure \ref{fig:knowndebris}b. To properly extract the belt location and detecting asymmetry, we use a resolvability of 3 as the criterion. }
    \label{fig:sb}
\end{figure*}

To illustrate, we assume all disks are belt-like with the peak radius inferred from the above method and a 20\% belt width (i.e., $\Delta r/r \sim$0.2). We can then compute the minimum S.B. assuming the disk is face-on at three selected wavelengths: 25 $\mu$m (MIRI/{\it JWST}), 50 $\mu$m (FIP/{\it Origins}) and 1.3 mm (Band6/ALMA). In order to probe the structures induced by unseen planets, the degree of resolvability (how well the structure is resolved) is another factor. We define the "resolvability" as the ratio between the diameter of the disk and the beam (FWHM) size. Fig. \ref{fig:multiwav_resolution}b shows two different resolvabilities by 3 and 10 beam sizes. The disk images resolved by $\gtrsim$3 beam sizes can pin point the location of the planetesimal belt and detect modest asymmetry while the ones resolved by 10 beam sizes are very diagnostic to resonance structures that are sensitive to the mass and location of a (unseen) perturbing planet. As shown in Fig.\ \ref{fig:knowndebris}, the angular sizes of the disks range from 0\farcs1 to a few 10\arcsec. We simply adopt a uniform beam size of 0\farcs82 to illustrate the expected S.B. for the debris sample and show the results in Fig.\ \ref{fig:sb}. These panels can be easily scaled to a different beam size, but the overall trend -- the higher the angular resolution the lower the surface brightness -- stands. To properly extract the belt location and detect asymmetry, we use a resolvability of 3 as the criterion for resolving disks, and assess the fraction of the disks that can be resolved by a given facility. A higher angular resolution is often achievable by changing the observed wavelength or array configuration. For example, ngVLA would provide milliarcsec resolution at 3 mm, enabling many new discoveries. However, its high angular resolution would over resolve the majority of the known disks, not efficient for studying a large sample. Our goal is to resolve the disks (by more than 3 resolution elements) at multiple wavelengths and assess the current and future prospective at the adapted resolution; therefore, the fraction of the resolved disks reported below is only comparable within the assumed parameters. 

\vspace{-2mm}
\vspace{-2mm}

\noindent \textit{\textbf{Millimeter Facility: }} The long baseline and large number of antennas provided by ALMA are producing many astonishing disk images; however, it would be difficult to resolve a large number of debris disks due to their faintness at millimeter wavelengths. At 1.3 mm, the expected continuum sensitivities are 10 and 5 $\mu$Jy/beam with an on-source integration time of 2.8 and 10 hours, respectively (see the two dashed lines in the left panel of Fig.\ \ref{fig:sb}). Assuming that ALMA can access 60\% of the sky, $\sim$8\% ($\sim$12\%) of the disks can be resolved with ALMA at S/N=10 with an on-source integration of 2.8 (10) hours. Probing debris in the terrestrial zone (warm belts) requires sub-arcsec resolution. For warm belts, limited by sensitivity, only $\sim$1\% are resolvable by ALMA and each with a 10-hour on-source integration.  Larger bandwidths, better receiver sensitivity and/or more antennas, highlighted in the ALMA Development Roadmap (Carpenter et al.\ 2019), might double the number of the resolved cold belts from ALMA in the coming decades. Nevertheless, probing warm debris for a large sample of systems would remain challenging with ALMA. 

\noindent \textit{\textbf{Mid-Infrared Facility: }}  JWST is our prime mid-infrared facility in the coming decade for resolving debris disks. Using MIRI in the F2550W filter, the pre-launched sensitivity is 7 $\mu$Jy/beam for a S/N of 10 in 2.8 hours. This sensitivity suggests that $\sim$45\% of the debris systems (in the white area of Fig.\ \ref{fig:sb}) can be easily resolved (at $\gtrsim$ 3 beam sizes) by JWST at 25 $\mu$m in terms of raw sensitivity. For systems that are relatively close with bright photospheric signals, it will be necessary to obtain point-spread-function reference images and/or utilize the coronographic mode to suppress the noise from stars. Although roughly half of the known debris disks can be resolved by JWST, only 2\% of them are warm belts (red dots in Fig.\ \ref{fig:sb}) which have small angular sizes. A space telescope that is larger than {\it JWST} or a space interferometer with capabilities enabling high angular resolution and sensitivity observations is needed to resolve the majority of the warm belts.

\noindent \textit{\textbf{Far-Infrared Facility: }} 
The cold debris has emission peaked at far-infrared, making actively-cooled space telescopes the primary choice for resolving cold belts. {\it Spica}, a proposed joint ESA/JAXA mission, is one of the candidates.  However, its 2.5 m aperture (smaller than {\it Herschel}) can only resolve nearby disks and its superb sensitivity would be hindered by background confusion. The 9-m {\it Origins} Space Telescope is the only far-infrared facility currently under study by NASA that can provide $\sim$1\arcsec\ resolution at far-infrared. A 9-m {\it Origins} would resolve the same debris sample as {\it JWST} (45\% of the debris disks), providing multi-wavelength, high-fidelity disk images. {\it Origins's} baseline mission ({\it JWST} size) would have a resolution of 2\farcs5, and is capable of resolving 20\% of the known disks (a factor of 3 improvement compared to {\it Herschel}); however, only a handful of the warm belts can be marginally resolved.  

\noindent \textit{\textbf{Final Remarks: }} In the coming decades, we would finally have the ability to resolve a large sample of debris disks and fully explore the potential of using them to probe the planets hidden from our detection techniques.  We also stress the need to resolve disks at multiple wavelengths to fully trace different grain populations, particularly in the millimeters. Therefore, the few systems that are close enough to be well resolved across the full wavelength range considered here provide the crucial foundation to interpret debris disk behavior in terms of the underlying planetary configuration. 

\section{Reference}
\vspace{-2mm}
Ack et al., 2012, A\&A, 540, 125 \\
Carpenter et al., 2019, arXiv:1902.02856  \\ 
Ertel et al., 2012, A\&, 544, 61 \\ 
Hughes et al., 2018, ARA\&A, 56, 541 \\
Krivov, 2010, Research in Astronomy and Astrophysics, 10, 383 \\
Raymond et al., 2012, A\&A, 541, 11 \\
MacGregor et al., 2017, ApJ, 823, 79 \\ 
MacGregor et al., 2017, ApJ, 842, 8 \\
Matra et al., 2018, ApJ, 859, 72 \\ 
Matthews et al., 2014,Protostars and Planets VI, 521 \\ 
Pawellek et al., 2014, ApJ, 792, 65 \\ 
Reach et al., 2010, Icarus, 209, 276 \\ 
Stenborg et al., 2018, ApJ, 868, 74 \\ 
Stapelfeldt et al., 2004, ApJS, 154, 458 \\
Su et al., 2009, ApJ, 705, 314 \\
Su et al., 2013, ApJ, 763, 118 \\
Wyatt 2008, ARA\&A, 46, 339 \\ 
Wyatt 2006, ApJ, 639, 1153 \\

\section{Co-Signers} 
\vspace{-2mm}

\begin{tabular}{lll}
\hline
Name    & Affiliation   &  E-mail   \\ 
\hline
Peter Plavchan	&	George Mason University	&	pplavcha@gmu.edu   \\
Christine Chen	&	Space Telescope Science Institute	&	cchen@stsci.edu	\\
Quentin Kral	&	LESIA	\&	 Paris Observatory	&	quentin.kral@obspm.fr	\\
Luca Matra	&	CfA	\&	 Harvard \& Smithsonian	&	luca.matra@cfa.harvard.edu	\\	
Michael Werner	&	JPL/Caltech	&	michael.w.werner@jpl.nasa.gov	\\	
Mark Booth	&	Friedrich-Schiller University	\&	 Jena	&	mark.booth@uni-jena.de	\\
John Debes  &	STScI	& debes@stsci.edu \\ 
\hline
\end{tabular}

\end{document}